\documentclass[twocolumn,prl]{revtex4}
\usepackage{graphicx}%
\usepackage{dcolumn}
\usepackage{amsmath}
\usepackage{latexsym}

\begin{document}
\title{Constrained deformation of a confined solid: a strain induced 
crystal-smectic transition} 
\author{Debasish Chaudhuri and Surajit Sengupta\\
}
\affiliation{
Satyendra Nath Bose National Centre for Basic Sciences, \\
Block-JD, Sector-III, Salt Lake,
Calcutta - 700098.
}
\date{\today}
\begin{abstract}
We report results of computer simulations of two-dimensional hard disks 
confined within a quasi one-dimensional ``hard-wall'' channel, a few atomic 
radii wide. Starting from a commensurate triangular solid a rescaling of the 
system size parallel to the channel length introduces a rectangular 
distortion of the solid which, beyond a critical limit, phase separates   
into alternating bands of solid and {\em smectic} phases.  
The resulting solid- smectic interfaces are broad and 
incorporate misfit dislocations. The stress-strain curve 
shows large plastic deformation accompanying  
the crystal-smectic transition which is reversible. The smectic 
phase eventually melts into a modulated liquid with a 
divergent Lindemann parameter. 
\end{abstract}
\maketitle

\noindent
Studies of small assemblages of molecules with one or 
more dimensions comparable to a few atomic spacings are 
significant in the context of nano-technology\cite{nanostuff}. 
Designing nano-sized machines requires a knowledge of the mechanical 
behavior of systems up to atomic scales, where, a priori, there is no 
reason for continuum elasticity theory to be valid. In most cases, however, 
the effects of finite size are relatively mild, showing up mainly as a 
variation of the numerical value of the elastic constants\cite{micrela} as a 
function of length scale. In this Letter, we show, on the other hand,
that small size and hard constraints can produce essentially new phenomena 
without a counterpart in the bulk system. 
We perform computer simulations of the simplest possible, 
nontrivial, molecular system, namely, two-dimensional hard disk ``atoms'' 
confined within a quasi one-dimensional channel; the physics of which is 
entirely governed by geometry. Bulk hard disks in two dimensions are known 
to melt\cite{alzowe,jaster,srnb} from a high 
density triangular lattice to an isotropic liquid with a narrow 
intervening hexatic phase\cite{kthny,srnb}. In contrast, 
for channel widths of a few atomic spacings, we find  
evidence for a {\em smectic} phase which nucleates 
as prominent  bands within the solid. 
The smectic phase arises when the size of the system 
in the direction parallel to the fixed walls is increased. A crystal to 
smectic transition, though predicted for anisotropic molecules\cite{smectic1}
is unusual for hard disks -- the anisotropy in this case arising 
purely from the external confining potential. In this respect, our 
results resemble the phenomenon of laser induced freezing\cite{life} where 
an external modulated electric
field produced by crossed  laser beams induces a series of phase
transitions\cite{lif2,lif3} involving triangular solid, 
modulated liquid as well as smectic phases. The analog of the modulating 
potential in our  case is, of course, provided by the walls which induce a 
periodic potential of mean force\cite{hansen-macdonald, hartmut} decaying 
with distance from the walls. The nature of the crystal- smectic transition
in our system, as we show below, are, however, different. 
Our results may be directly verified in experiments on sterically stabilized
``hard sphere'' colloids\cite{colbook} confined in glass channels 
and may also be relevant for similarly confined atomic systems interacting 
with more complex potentials.

\begin{figure}[t]
\begin{center}
\includegraphics[width=8.6cm]{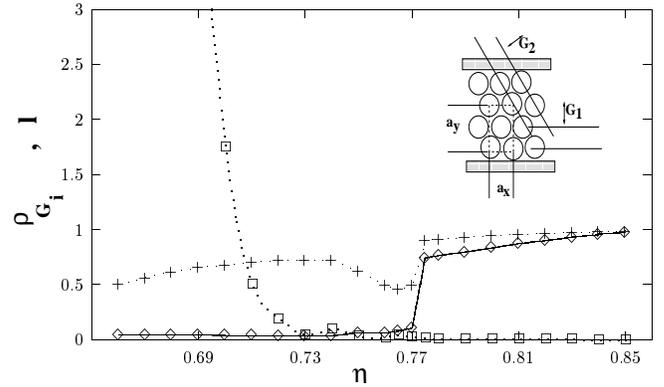}
\end{center}
\caption{Results of NVT ensemble Monte Carlo simulations of 
$N = n_x \times n_y = 65 \times 10$ 
hard disks confined between two parallel hard walls separated by a distance
$L_y = 9.001\, {\rm d}$ where ${\rm d}$ is the hard disk diameter. Inset shows 
the geometry used; only four layers have been shown for clarity.
The reciprocal lattice vectors (RLVs)$\,{\bf G_1}$ and ${\bf G_2}$, 
the rectangular unit cell and the lattice parameters $a_x$ and $a_y$ are 
also shown. At $\eta = 0.85$ we have a strain free triangular lattice.
Plots show $\rho_{\bf G_i}, i = 1 (+),2(\diamond)$
the structure factor for RLVs ${\bf G_i}(\eta)$, averaged over 
symmetry related directions, as a function of $\eta$. The 
value of $\eta$ is changed by changing the length of the box $L_x$ while keeping
$L_y$ and $N$ fixed.  
For each $\eta$, the 
system was equilibrated over $10^6$ Monte Carlo steps (MCS) and data averaged over a further
$10^6$ MCS. $\rho_{\bf G_1}$ is non-zero throughout implying long 
ranged orientational order. 
In contrast, $\rho_{\bf G_2}$ jumps to zero at $\eta = \eta_{c_1}\approx .77$. 
Also plotted in the same graph is the 
Lindemann parameter $l(\Box)$ which diverges below  
$\eta = \eta_{c_3} =  .7 < \eta_{c_1}$. 
Note that at $\eta_{c_1}$ 
$\chi = 9.54 \approx n_l(=10)-1/2$.
The lines in the figure are a guide to the eye.}
\label{order}
\end{figure}
\vskip .2cm

\noindent
The bulk system of hard disks where particles $i$ and $j$, in two dimensions, 
interact with the potential $V_{ij} = 0$ for $|{\bf r}_{ij}| > {\rm d}$ and 
$V_{ij} = \infty$ for $|{\bf r}_{ij}| \leq {\rm d}$, where ${\rm d}$ is 
the hard disk diameter and ${\bf r}_{ij} = {\bf r}_j - {\bf r}_i$ the 
relative position vector 
of the particles, has been extensively\cite{alzowe,jaster,srnb} 
studied. Apart from being easily 
accessible to theoretical treatment\cite{hansen-macdonald}, experimental systems
with nearly ``hard'' interactions viz. sterically stabilized
colloids\cite{colbook} are 
available. The hard disk free energy is entirely entropic in 
origin and the only thermodynamically relevant variable is the number density   
$\rho = N/V$ or the packing fraction $\eta = (\pi/4) \rho {\rm d}^2$. 
Simulations\cite{jaster}, experimental\cite{colbook} and 
theoretical\cite{rhyzov} studies of hard 
disks show that for $\eta > \eta_f = .719$ the system exists as a triangular 
lattice which transforms to a liquid below $\eta_m = .706$. The small 
intervening region contains a hexatic phase predicted by 
the Kosterlitz-Thouless-Halperin-Nelson-Young theory\cite{kthny}
of two dimensional melting.
The surface free energy of the hard disk system in contact with a hard wall
has also been obtained\cite{hartmut} taking care that the 
dimensions of the system are compatible with a perfect (strain-free) 
triangular lattice. The effect of in-commensuration, though, which may routinely 
arise in experiments, surprisingly, have not received the attention 
it deserves. We attempt to address this aspect as follows. 
\begin{figure}[t]
\begin{center}
\includegraphics[width=8.6cm]{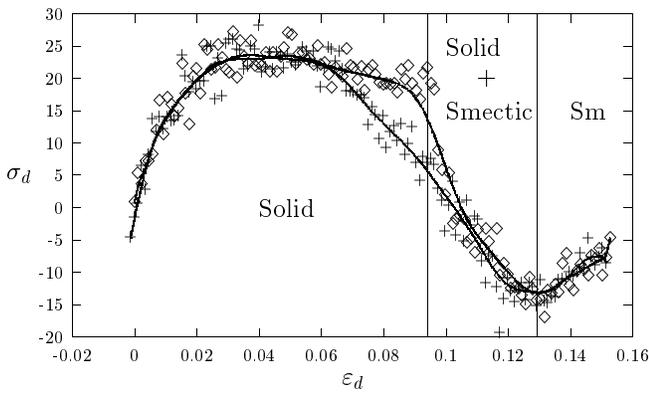}
\end{center}
\caption{ A plot of the normal stress $\sigma_d$ 
versus the conjugate strain $\varepsilon_d =(\eta_0 - \eta)/\eta_0 $
($\eta_0 = 0.85$) obtained from our Monte Carlo simulations of $65 \times 10$
hard disks showing a typical Van der Waals loop in the constant strain 
ensemble.  Data for the plot is obtained by 
equilibrating at each strain value for $2\times10^4$ MCS and averaging the 
data for a further $3\times10^4$ MCS. 
The stress for the hard disk system has been calculated by the 
standard method\cite{elast} by averaging the collision probability. It is minimum at $\eta=\eta_{c_2}\approx .74$.
The entire cycle consisting of increasing 
$\varepsilon_d (\diamond)$ and again decreasing to zero $(+)$ 
using typical parameters appropriate for an atomic system, corresponds to a 
real frequency of $\approx 100 {\rm K Hz}$. The lines in the figure are a
guide to eye. The solid ($\eta > \eta_{c_1}$), two phase ($ \eta_{c_2} < 
\eta < \eta_{c_1}$) and smectic (Sm) ($\eta < \eta_{c_2}$) regions are 
indicated in the figure. We have repeated this 
calculation with a cycle frequency $10 {\rm K\,Hz}\,- 1{\rm M\,Hz}$ with 
no essential change in the results. 
}
\label{stress}
\end{figure}
\vskip 0.2cm

\noindent
Consider a narrow channel in two dimensions  of width $L_y$ defined by 
hard walls at $y = 0$ and $L_y$ 
($V_{\rm wall}({\bf r}) = 0$ for $ 0 < r_y < L_y$ and $ = \infty$ otherwise)
and length $L_x$ with $L_x \gg L_y$. Periodic boundary 
conditions are assumed in the direction $x$ implying $x + L_x = x$. 
In order that the channel may accommodate
$n_{l}$ layers of a homogeneous, triangular lattice with lattice parameter
$a_0$ of hard disks of diameter ${\rm d}$, (Fig.\ref{order}) one 
needs, 
\begin{equation}
L_y = \frac{\sqrt{3}}{2}(n_{l} - 1) a_0 + {\rm d} 
\label{perfect}
\end{equation}

\noindent 
Defining $\chi = 1 + 2(L_y - {\rm d})/\sqrt{3} a_0$,
the above condition reads $\chi = {\rm integer} = n_{l}$ and violation
of Eqn.(\ref{perfect}) implies a rectangular strain away from the reference
triangular lattice of $n_l$ layers. The lattice parameters of a centered 
rectangular (CR) unit cell are $a_x$ and $a_y$ (Fig. \ref{order} inset). In 
general, for a CR lattice with given  $L_y$ we have, 
$a_y = 2 (L_y - {\rm d})/(n_l-1)$ and, ignoring vacancies,
$a_x = 2/\rho a_y$. The normal strain $\varepsilon_d = 
\varepsilon_{xx} - \varepsilon_{yy}$ is then,
\begin{equation}
\varepsilon_d =  \frac{n_l - 1}{\chi - 1} - \frac{\chi - 1}{n_l - 1},
\label{strain}
\end{equation}

\noindent
where the number of layers $n_l$ is the nearest integer to $\chi$ so that 
$\varepsilon_d$ has a discontinuity
at half~-integral values of $\chi$. For large $L_y$ this discontinuity and 
$\varepsilon_d$ itself vanishes as $1/L_y$ for all $\eta$. 
\vskip .2 cm

\noindent
We study the effects of this strain $\varepsilon_d$ on the 
hard disk triangular solid at fixed $L_y$ large enough to accommodate a small 
number of layers $n_l \sim 9 - 25$. The strain $\varepsilon_d$ is
imposed by expanding the dimension of the system $L_x$ parallel to the walls 
keeping $L_y$ fixed so that $\varepsilon_d = (\eta_0 - \eta)/\eta_0$, where 
$\eta_0$ is the packing fraction corresponding to an unstrained triangular
solid. We monitor the Lindemann parameter 
$l = < ({u^x}_i - {u^x}_j)^2>/a_x^2 + < ({u^y}_i - {u^y}_j)^2>/a_y^2 $ 
where the angular brackets denote averages over configurations, 
$i$ and $j$ are nearest neighbors and ${u^{\alpha}}_i$ is the $\alpha$-th 
component of the displacement of particle $i$ from it's mean position. 
The parameter $l$ diverges at the melting transition \cite{Zahn}.
We also measure the structure factor 
$\rho_{\bf G} =\left| \left< \frac{1}{N^2} \sum_{i,j = 1}^N 
\exp(-i {\bf G}.{\bf r}_{ij})\right> \right|,$ 
for ${\bf G} = \pm {\bf G_1}(\eta)$, the 
reciprocal lattice vector (RLV) corresponding to the set of close-packed 
lattice planes of the CR lattice perpendicular to the 
wall, and ${\bf \pm G_2}(\eta)$ the 
four equivalent RLVs for close-packed planes at an angle 
( $ = \,\pi/3$ and $2\pi/3$ in the triangular lattice) to the wall
(see Fig. \ref{order} inset). 
\vskip .2cm

\noindent
We find, throughout, $\rho_{\bf G_2} <  \rho_{\bf G_1} \neq 0$, a 
consequence 
of the hard wall constraint\cite{hartmut} which manifests as an oblate 
anisotropy 
of the local density peaks in the solid. 
As $\eta$ is decreased (see Fig. \ref{order} for details) both 
$\rho_{\bf G_1}$ and $\rho_{\bf G_2}$ show 
a jump at $\eta = \eta_{c_1}$ close to $\chi \approx n_l - 1/2$. 
For $\eta < \eta_{c_1}$ we get  
$\rho_{\bf G_2} = 0$ with $\rho_{\bf G_1} \not= 0$ signifying 
a transition from crystalline to smectic like  order. 
The Lindemann parameter $l$ remains zero and shows a divergence only below 
$\eta = \eta_{c_3}(\approx \eta_m)$ indicating a finite-size-
broadened melting of the smectic to a modulated liquid phase. 
We have also calculated the normal stress 
$\sigma_d = \sigma_{xx} - \sigma_{yy}$ (see Fig. \ref{stress}). 
For $\eta = \eta_0$ the stress is 
purely hydrostatic with $\sigma_{xx} = \sigma_{yy}$ as expected. As
$\eta$ decreases, the 
stress increases linearly in the elastic limit, flattening out at the  
onset of non-linear behavior at $\eta \stackrel{<}{\sim} \eta_{c_1}$. 
At $\eta_{c_1}$, $\,\,\sigma_d$ decreases and eventually becomes negative. 
On further decrease in $\eta$ below $\eta_{c_2}$ (Fig. \ref{stress}),$\,\,\sigma_d$ approaches 
$0$ from below thus forming a Van der Waals loop typical of the constant 
strain ensemble. If the strain is reversed by increasing $\eta$
back to $\eta_0$ the entire stress-strain curve is traced back 
with no remnant stress at $\eta = \eta_0$ showing that the 
plastic region is reversible. As $L_y$ is increased,
$\eta_{c_1}$ merges with $\eta_{c_3}$ for $\chi \stackrel{>}{\sim} 25$.
If instead, $L_x$ and $L_y$ are both rescaled to keep $\chi$ fixed or 
periodic boundary conditions are imposed in both $x$ and $y$ directions, the 
transitions in the various quantities occur approximately simultaneously
as expected in the bulk system. Varying $n_x$ in the range $10 - 1000$ produces
no essential change in results.
\vskip .2 cm

\begin{figure}[t]
\begin{center}
\includegraphics[width=7.0cm]{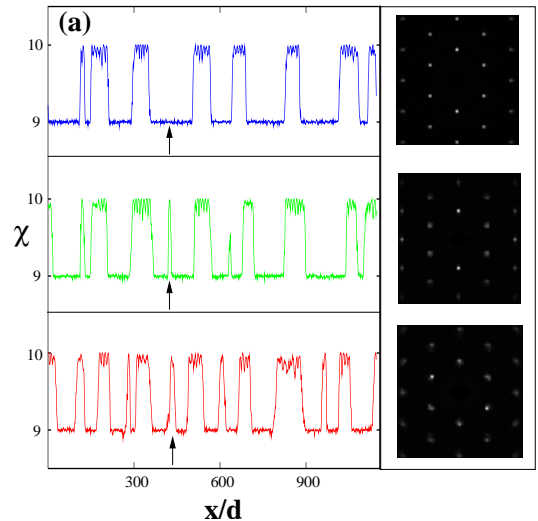}
\vskip 0.2cm
\end{center}
\caption{(color on-line)
(a) Plot
of $\chi(x)$ as a function of the 
$x/{\rm d}$ at $\eta = .76$ 
after $3\times10^5$, $5\times10^5$ and $2\times10^6$ (top) MCS for 
$N = 10^4$. Note that $\chi = 10$ in the solid and $ = 9$ in the smectic 
regions. Arrows show the coalescence of two bands as a function of time.
The panel on the right shows calculated diffraction patterns for 
the solid (top), smectic (middle) and inter-facial (bottom) regions.
Inversion symmetry is broken at the interface because of the 
inter-facial dislocation (see below).
(b) Close up view of a crystal-smectic  interface from superimposed 
positions of $10^3$ configurations at $\eta = .77$. The colors 
code the local density of points from red (high) to blue (low). 
Note the misfit dislocation in the inter-facial region.
}
\label{interface}
\end{figure}

\noindent
For $\eta_{c_3} < \eta < \eta_{c_1}$ we observe that the smectic  
order appears within narrow bands (Fig. \ref{interface}) which nucleate 
at $\eta_{c_1}$. Inside these bands the number of layers is less 
by one and the system in this range of $\eta$ is in a mixed phase. 
A plot (Fig.\ref{interface}(a))
of $\chi(x)$, obtained by averaging the instantaneous $a_y$ from particle 
configurations over a strip spanning $L_y$, shows 
bands in which $\chi$ is less by one compared to the 
crystalline regions. Once nucleated narrow bands coalesce 
to form wider bands, the dynamics of which is, however, extremely slow. 
The bands grow as $\eta$ is decreased. Calculated diffraction 
patterns (Fig. \ref{interface} (a)) show that, locally, within a 
smectic band $\rho_{\bf G_1} \gg  \rho_{\bf G_2}$ in contrast to the solid 
region where $\rho_{\bf G_1} \approx \rho_{\bf G_2} \neq 0 $. 
\vskip .2 cm

\noindent
Strong finite size corrections makes a complete theoretical treatment of this 
problem difficult. However, significant progress may be made using qualitative
arguments as we show below. The total free energy of the system ${\cal F}^T$ 
may be decomposed as,
\begin{equation}
{\cal F}^T(\eta,\chi) = K^{\Delta}(\eta)\varepsilon_d^2(\chi) + {\cal F}^{\Delta}(\eta)
\label{totf}
\end{equation}
where $K^{\Delta}(\eta)$ is an elastic constant and ${\cal F}^{\Delta}(\eta)$ 
the free energy of the perfect triangular lattice in contact with a 
hard wall\cite{hartmut} at packing fraction $\eta$. It is clear that 
${\cal F}^T$ has minima for all $\chi= n_l$. For half integral values 
of $\chi$ the crystalline structure is metastable with respect to an 
intervening smectic when adjacent local density peaks of the 
solid overlap in the $x$ direction. 
This overlap is facilitated by the (oblate) anisotropy of the solid 
density peaks. For large $L_y$ the minima in 
${\cal F}^{T}$ merge to 
produce a smooth free energy surface independent of $\chi$.
For small $L_y$ all regions of the parameter space corresponding to   
non-integral $\chi$ are also {\em globally} unstable as belied by the loop 
in the stress-strain curve (Fig.\ref{stress}). The system should therefore 
break up into 
regions with various $n_l$.  Such fluctuations are, however, suppressed due to  
the structure of interfaces between regions of differing $n_l$. A 
superposition of 
many particle positions near such an interface (see Fig. \ref{interface}(b)) 
shows that: 
$(1)$ the width of the interface is large, spanning about $10 - 15$ atomic 
spacings and $(2)$ the interface between $n_l$ layered crystal and $n_l -1$ 
layered smectic contains a {\em dislocation} with Burger's vector in the $y$ 
direction which makes up for the difference in the number of layers. 
Note that the presence of these dislocations breaks inversion symmetry as 
observed in the local diffraction patterns. 
The core region of this dislocation extends over 
many layers with some of the disks within the interface alternating  
between positions corresponding to either a smectic or a solid. 
Each band of width $s$ is therefore held in place by a 
dislocation-anti-dislocation pair (Fig. \ref{interface}). 
In analogy with classical nucleation theory\cite{cnt}, the 
free energy $F_b$ of a single band can be written as 
\begin{equation}
 F_b = -\Delta F s + E_c + \frac{1}{4\pi}b^2 K^\Delta \log \frac{s}{a_0}  
\label{becker-doring}
\end{equation}
where $b = a_y/2$ is the Burger's vector,  
$\Delta F$ the free energy difference between the crystal 
and the smectic per unit length and $E_c$ the core energy for 
a dislocation pair.  
Bands nucleate when dislocation pairs separated by 
$s > \frac{1}{4\pi}b^2 K^\Delta/\Delta F$
arise due to random fluctuations. Band coalescence occurs by diffusion 
aided dislocation ``climb'' which is extremely improbable 
in a high density phase leading to slow kinetics. 
The growing smectic bands are in a state of tension in the $y$ direction
implying $\sigma_{yy} > \sigma_{xx}$ which is countered by the 
compressive stress in the crystalline region. 
Growth of smectic bands therefore reduces $\sigma_d$ which attains a 
minimum at $\eta = \eta_{c_2}$ when a single band spans the entire length. 
Subsequently $\sigma_d \to 0$, the value in the liquid phase.
Since orientation relationships between the 
crystal and smectic are preserved, the stress-strain relationship is 
completely determined by
the amount of the co-existing phases which   
explains the reversibility\cite{onions}. 
For large values of $L_y$ the smectic phase vanishes since the 
strains involved themselves go to zero. Nevertheless, transition between 
$n_l$ and $n_l \pm 1$ layered crystals have been observed by us. 
We must mention here that the choice of the ensemble, viz. constant 
strain, is crucial since, in the constant stress ensemble, the hard disk 
system fails at $\eta_{c_1}$ and the strain diverges producing 
a homogeneous low density gas with no interface. 
Finally, our smectic bands are
reminiscent of ``slip'' or ``deformation'' bands which arise during plastic
flow of macroscopic ductile materials\cite{cahn-haasen}.
\vskip .2 cm

\noindent
Apart from constrained hard sphere colloids\cite{colbook} where our results 
are directly testable, strain induced crystal-smectic transition may be 
observable in experiments on the deformation of mono-layer atomic nano-beams 
or strips of real materials confined to lie within a channel\cite{nanostuff}. 
This is because the constraints we choose to study are geometrical and would 
exist in any system. The occurrence of a smectic phase 
may be important for the tribological\cite{tribo} properties of nano-scale
machine parts where atomic friction plays an important role in their function.
In the future we would like to study the kinetics of the crystal-
smectic transition in detail as well as the effect of substrate disorder
on the nature of this transition.
\vskip .2 cm

\noindent
The authors thank M. Rao, V. B. Shenoy and A. Datta 
for useful discussions; D. C. thanks C.S.I.R., India, for a fellowship. 
Financial support by Department of Science and Technology, 
India, is gratefully acknowledged.

 

\end{document}